# Who Wins the Game of Thrones? How Sentiments Improve the Prediction of Candidate Choice

Chaehan So
Information & Interaction Design
Humanities, Arts & Social Sciences Division, Yonsei University
Seoul, South Korea
Email: cso@yonsei.ac.kr

*Abstract*— This paper analyzes how candidate choice prediction improves by different psychological predictors. To investigate this question, it collected an original survey dataset featuring the popular TV series "Game of Thrones".
The respondents answered which character they anticipated to win in the final episode of the series, and explained their choice of the final candidate in free text from which sentiments were extracted. These sentiments were compared to feature sets derived from candidate likeability and candidate personality ratings.
In our benchmarking of 10-fold cross-validation in 100 repetitions, all feature sets except the likeability ratings yielded a 10-11% improvement in accuracy on the holdout set over the base model. Treating the class imbalance with synthetic minority oversampling (SMOTE) increased holdout set performance by 20-34% but surprisingly not testing set performance.
Taken together, our study provides a quantified estimation of the additional predictive value of psychological predictors. Likeability ratings were clearly outperformed by the feature sets based on personality, emotional valence, and basic emotions.

*Keyword*s—candidate prediction, sentiment analysis, likeability, emotional valence, basic emotions, multi-class imbalance.

I. INTRODUCTION

In political elections, people vote for their favorite candidate based on a variety of criteria. The three criteria most influencing voters' choice are the candidate's perceived competence, trustworthiness, and warmth or likeability - all well investigated in political election research [1]. Apart from these criteria, however, people also base their decision on the political party the candidate belongs to. Research in this domain largely ignores the effect of party-loyalty, party preference, or preference for a political agenda that confounds the candidate choice. Other influences interacting with the perception of candidate traits are trait associations at the party brand level, specifically on political party brand image, advertising spending and negative advertising [2]. Taken together, multiple factors cause bias in people's political candidate choice. Finding a context outside a political election setting would therefore enable to investigate the pure effect of psychological predictors on candidate choice. Therefore, the present work aims to contribute to the literature in two aspects:

First, this work created a context of people voting for their favorite candidate that highly resembles a political election but is void of any political context that confounds people's candidate choice. To accomplish this goal, this work solicited people who had watched the popular TV series "Game of Thrones" (GOT) to predict the character who would win the game of thrones, i.e. who would become the king or queen of all kingdoms in the final episode of the last series aired on 19 May, 2019.

Second, this work disentangles psychological predictors into separate categories, and compares their prediction quality. The resulting benchmarking of predictor types allows to design electoral studies based on the effectiveness of predictors types, and thus make election surveys more efficient. To reach this goal, the current work derived several predictor types from election research in political psychology, namely demographic variables, likeability ratings and sentiment analysis. These predictor types will be delineated in the following.

*A. Likeability*

People's judgment of someone's likeability is essential for a positive first impression. Likeability is the perception of warmth that is associated with caring and sociability of the target person [2]. Its relevance was highlighted in the US presidential campaign 2008 when presidential candidate Barack Obama was consistently rated more likeable than Hillary Clinton, which was seen as the reason for his higher poll ratings [3].

*B. Personality*

Psychological research found that personality is a multi-faceted construct that consists of several separable dimensions called traits [4]. The most commonly used personality trait model is the Five-Factor Model (FFM) of personality introduced by McCrae and John [5], referenced as the *big five* personality traits: openness, conscientiousness, extraversion, agreeableness and neuroticism. The perception of candidates' personality traits have been shown to influence voters' choices in election research [2], [6].

*C. Sentiment Analysis*

Sentiment analysis gained popularity among machine learning researchers in the last decade as an effective tool of

NLP (Natural Language Processing) to analyze qualitative data by retrieving the emotional content of user data. Sentiment analysis has been applied to predict the evolution of the stock market [7], or to predict commercial success based on sentiments from hotel reviews [8], movie reviews [9], or restaurant reviews [9], or to predict risk phenomena, e.g. crime [10] or epidemic outbreaks [11].

A major portion of sentiment analysis research has focused on predicting election outcomes based on Twitter sentiments, of e.g. the US presidential elections [12], the national parliament election in Germany 2009 [13], the general elections in Pakistan 2013 [14] or the presidential election in Brazil 2014 [15]. The methodological approaches of this research differ in the unit of text analyzed, varying from word, sentence, paragraph [15], to a semantic unit such as aspect [16], opinion [17], topic [18], or political issue [19].

*D. Sentiments: Emotional Valence & Basic Emotions*

Psychological research investigated how emotion influences judgment and found that a key factor is *emotional valence*, i.e. the distinction whether the emotion is positive or negative [20]. There is converging evidence from neuroscientific research on the crucial role of emotional valence on cognitive processing, evidenced by event-related potentials (ERPs) [21] and by fMRI experiments [22]. In sentiment analysis, the view that emotional valence is a continuum between negative to positive was replaced by the notion that both positive and negative sentiments can coincide, and, except for extreme levels, are largely independent [23].

Another sentiment category is *basic emotions*. The theory of basic emotions by Ekman [24] postulates that certain emotions are universal as they are biologically based and exist across cultures. Neuroimaging research found evidence for the six basic emotions initially suggested by Ekman, namely happiness, surprise, fear, sadness, anger, disgust [25], [26].

## II. METHOD

*A. Participants*

The empirical study used Amazon Mturk to acquire a sample of 1279 participants in the last week (13-19 May 2019) before the last Game of Thrones episode aired on television. For the data collection, an online questionnaire was answered by 1279 participants. 168 survey responses had empty or meaningless answers (e.g. "yes", "like") for the free text field explanation of their candidate choice (throne_why) and were thus removed, leading to a final sample size of n = 1111.

*B. Descriptive Statistics*

Participants were 56.2% male and 43.1% female, and on average 30.58 (SD = 8.30) years old. Their highest attained education level was 57.2 % *Bachelor's degree*, followed by 18.2% *Master's degree*, 17.7% *High School*, 5.0% *Professional degree*, 1.0% *not finished High School*, and 0.9% *Ph.D. degree*. Ethnicity was spread over *White* (47.5%), *Asian or Pacific Islander* (36.6%), *Black or African American* (5.22%), *Hispanic or Latino* (5.0%), *Mixed* (1.5%), and *Middle Eastern* (1.2%). 14.9% of the participants rated their expertise level as *extremely high*, 34.3% as *very high*, 25.4% as *quite high*, 16.1% as *moderately high*, 6.4% as *somewhat high*, 2.34% as *not high* and 0.6% as *not high at all*.

*C. Survey Design*

The survey was constructed similar to a poll of favorite candidate selection with ratings of likeability and personality traits for each potential candidate (psychological predictors), the candidate choice (throne) and a free text explanation to explain the candidate choice (throne_why). The latter was used to create new features (emotional valence and basic emotions) by sentiment analysis to find out whether and to what degree sentiment features improve the prediction of candidate choice.

*D. Data Preprocessing*

The goal of the data cleaning procedure was to remove any information that is directly predictive of the target variable, i.e. the candidate choice. Therefore, the free text variable throne_why was cleaned from any word occurrence of candidate names and their most frequent misspellings (e.g. "John Snow" instead of "Jon Snow" or "Daenarys" instead of "Daenerys").

Furthermore, 106 empty text responses and 62 responses containing a single word ("yes" or "like") were removed from the throne_why variable leading to an effective sample size of 1111 observations. These meaningless responses were probably given by Amazon Mturk participants who aim to finish the survey as quickly as possible.

*E. Feature Sets*

The survey data was divided into four feature sets displayed in Figure 1 and described in the following.

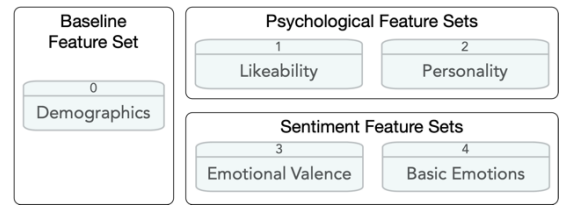

Figure 1: Feature Sets

*1) Baseline Feature Set*

From the descriptive statistics variables, a baseline feature set was created that consisted of the demographic variables age, expertise level, education and watch experience.

The watch experience was categorized into six levels (from 1 = not season 8 to 6 = all episodes of seasons 1-8 repeatedly), and the expertise level was answered on a 7-point Likert scale (from 1 = *Not high at all* to 7 = *Extremely high*). The expertise level was encoded on a 7-point Likert scale ranging from 1 = *Not high at all* to 7 = *Extremely high*. Likewise, watch experience was encoded on a 7-point Likert scale from 1 = *Not season 8 so far* to 6 = *All episodes of seasons 1-8 repeatedly*. The education level was encoded numerically from 1 = *Did not finish High School* to 6 = *Ph.D.*

*2) Likeability Feature Set*

The survey participants rated the degree to which they liked each of the twelve candidates (from 1 = *Like not at all* to 5 = *Like extremely*).

*3) Personality Feature Set*

The big five personality items were based on the 20 items of the Mini-IPIP questionnaire [27], a short version of the FFM inventory, and encoded on a 5-point Likert scale (from 1 = *Strongly Disagree* to 5 = *Strongly Agree*).

*4) Sentiment Feature Sets: Emotional Valence & Basic Emotions*

The sentiments feature sets were created by adding sentiments as predictors by a lexical approach. This study chose the NRC sentiment lexicon [28], [29] from the tidytext package [30] because it represents one of the largest available lexicon corpora with 13901 word-to-sentiments mappings. Compared to other sentiment lexica (afinn, bing, loughran), it further provides two types of sentiments – emotion valence (negative, positive) and eight basic emotions (anger, anticipation, disgust, fear, joy, sadness, surprise, trust). These two types were used to create separate feature sets of emotional valence and basic emotions.

Each free text response explaining candidate choice was mapped with the NRC sentiment lexicon. The sentiments counts were normalized per participant in order to accommodate higher sentiment counts which had been inflated by text length. These normalized sentiment scores were used as sentiment features.

*F. Benchmarking Method*

The present work chose two different implementations of the same algorithm where possible to detect implementation-specific tendencies: knn [31] and kknn [32] for k-nearest Neighbors, svmRadial and svmLinear [both 52] for Support Vector Machines, randomForest [34] and ranger [35] for Random Forests, xgbTree and xgbLinear [36] for XGBoost. In addition, the packages nnet [37] for single-layer neural networks and gbm [38] for Gradient Boosting Machines were selected.

All five feature sets (baseline, likeability, personality, emotional valence, basic emotions) were analyzed by the same model training of 100 repetitions of 10-fold cross-validation. The performance criteria accuracy and kappa were calculated by averaging across the 1000 folds on the hold-out validation set.

The majority class represented more than 40% of the target variable. To correct such class imbalance, an effective method is SMOTE, the synthetic minority oversampling technique [39]. The SMOTE algorithm detects the minority class and generates (synthesizes) new instances of this class by a knn interpolation. The additional instances serve to attenuate the bias introduced by the majority class.

*G. Technological Infrastructure*

All analyses were conducted with Rstudio Server on a virtual machine with Ubuntu 16.04 LTS, 96 CPU cores (Intel Skylake), and 250 GB RAM in the Google Cloud Compute Engine.

### III. RESULTS

The baseline feature set contained features that are not semantically related to any candidate preference. As expected, the prediction performance is rather weak, in a range between 28.6% - 37.9% accuracy on the testing set (see Table 1). As the cross-validation procedure estimates accuracy on the hold-out

TABLE 1: BENCHMARKING ACCURACY RESULTS – BASELINE FEATURE SET

| descriptive feature set | | | | | |
|---|---|---|---|---|---|
| algorithm | Cross-valid Acc – mean | Cross-valid Acc – sd | Cross-valid Kappa – mean | Cross-valid Kappa – sd | Testingset Accuracy |
| svmRadial | 37.7% | 2.1% | 6.8% | 3.2% | 37.9% |
| xgbTree | 37.8% | 2.4% | 8.5% | 3.4% | 37.9% |
| nnet | 37.5% | 2.7% | 9.9% | 3.7% | 37.3% |
| ranger | 36.0% | 2.8% | 9.5% | 4.0% | 37.3% |
| svmLinear | 36.1% | 1.3% | 2.7% | 3.5% | 37.3% |
| rf | 34.6% | 2.9% | 9.4% | 4.0% | 36.6% |
| gbm | 37.0% | 2.4% | 8.3% | 3.5% | 35.4% |
| knn | 31.2% | 3.4% | 6.7% | 4.4% | 35.4% |
| xgbLinear | 28.5% | 3.6% | 8.2% | 4.4% | 29.8% |
| kknn | 28.6% | 3.5% | 8.6% | 4.2% | 28.6% |

*Note: Accuracy on 10-fold cross-validation in 100 repetitions, train/test split 85:15*

set, and due to stratification in the data split, the accuracy does not strongly differ between training and testing set, as expected.

The result of the first experimental run benchmarking the original dataset are shown in Table 2. They show the four feature sets of psychological predictors (likeability, personality) and sentiment predictors (emotion valence, basic emotions).

The mean performance for the likeability feature set is surprisingly weak. The best model reaches only approximately the testing set accuracy of the baseline. It follows that the likeability predictors do not provide any substantial predictive value. In contrast, the personality, emotions-valence and basic-emotions feature sets perform distinctly better than the likeability feature set. They are very similar with 51-52% accuracy as best testing set performance.

The top algorithms in testing set performance are the random forests package *ranger* and neural networks package *nnet*, both ranked first twice. On second rank, the support vector machines with linear kernel *svmLinear* appears twice. Next, the gradient boosting machines package gbm appears on 2[nd] and 3[rd] rank.

The lowest performance is consistently shown by the k-Nearest Neighbors implementations (kknn, knn). The fact that the random forest implementations (ranger, rf) perform on the lowest ranks for the sentiment feature sets is surprising.

Overall, the psychological feature sets can be well predicted by random forests (ranger), whereas the sentiment feature sets seemed to be best predictable for a single-layer neural network (nnet) or support vector machines with linear kernel (svmLinear). It is interesting that the performance was very similar between the last three feature sets even though they distinctly differed in the number of predictors. That means that the two predictors in the emotions-valence feature set were as predictive as the 20 predictors of the personality feature set.

The surprising summary of these findings is that the likeability feature set showed very poor performance similar to the baseline, where the remaining feature sets showed an increase of 14.2-15.5% in testing set accuracy.

TABLE 2: BENCHMARKING ACCURACY RESULTS FOR ALL FEATURE SETS – UNBALANCED DATASET

| psychological – likeability feature set | | | | | | psychological – personality feature set | | | | | |
|---|---|---|---|---|---|---|---|---|---|---|---|
| algorithm | Cross-valid Acc – mean | Cross-valid Acc – sd | Cross-valid Kappa – mean | Cross-valid Kappa – sd | Testingset Accuracy | algorithm | Cross-valid Acc – mean | Cross-valid Acc – sd | Cross-valid Kappa – mean | Cross-valid Kappa – sd | Testingset Accuracy |
| ranger | 34.3% | 2.2% | 6.8% | 3.1% | 37.4% | ranger | 48.2% | 5.6% | 28.1% | 7.9% | 52.6% |
| rf | 34.2% | 2.6% | 8.5% | 3.5% | 36.4% | gbm | 47.5% | 6.0% | 30.0% | 7.9% | 51.3% |
| nnet | 35.0% | 2.5% | 12.0% | 3.2% | 35.8% | svmRadial | 47.2% | 3.9% | 18.4% | 6.5% | 51.3% |
| svmLinear | 34.1% | 1.7% | 2.7% | 2.3% | 35.8% | rf | 48.4% | 5.5% | 28.5% | 7.8% | 50.0% |
| svmRadial | 34.7% | 2.1% | 5.8% | 2.9% | 35.8% | xgbLinear | 47.0% | 5.8% | 30.3% | 7.5% | 50.0% |
| gbm | 34.3% | 2.5% | 7.9% | 3.5% | 35.3% | xgbTree | 48.6% | 5.4% | 27.7% | 7.7% | 48.7% |
| xgbTree | 34.4% | 2.4% | 7.1% | 3.3% | 34.2% | kknn | 41.1% | 5.9% | 23.4% | 7.4% | 47.4% |
| xgbLinear | 28.9% | 3.4% | 9.4% | 4.1% | 32.6% | svmLinear | 42.5% | 5.5% | 23.0% | 7.2% | 47.4% |
| knn | 29.8% | 3.2% | 8.3% | 3.9% | 31.0% | knn | 44.7% | 5.7% | 24.6% | 7.8% | 44.7% |
| kknn | 26.1% | 3.4% | 7.6% | 3.9% | 30.5% | nnet | 40.9% | 6.1% | 21.0% | 8.0% | 44.7% |
| sentiments – emotional valence – feature set | | | | | | sentiments – basic emotions – feature set | | | | | |
| algorithm | Cross-valid Acc – mean | Cross-valid Acc – sd | Cross-valid Kappa – mean | Cross-valid Kappa – sd | Testingset Accuracy | algorithm | Cross-valid Acc – mean | Cross-valid Acc – sd | Cross-valid Kappa – mean | Cross-valid Kappa – sd | Testingset Accuracy |
| nnet | 48.0% | 2.3% | 2.9% | 3.8% | 51.6% | nnet | 48.3% | 1.4% | 0.0% | 0.4% | 51.6% |
| svmLinear | 48.4% | 1.2% | 0.0% | 0.0% | 51.6% | svmLinear | 48.4% | 1.2% | 0.0% | 0.0% | 51.6% |
| gbm | 48.4% | 2.6% | 4.6% | 4.6% | 50.5% | svmRadial | 48.4% | 1.3% | 0.0% | 0.0% | 51.6% |
| xgbTree | 48.2% | 2.4% | 3.8% | 4.0% | 50.5% | gbm | 48.2% | 2.4% | 2.5% | 3.9% | 50.5% |
| svmRadial | 48.4% | 1.9% | 2.5% | 2.9% | 49.5% | xgbTree | 47.8% | 2.3% | 1.6% | 3.5% | 49.5% |
| rf | 43.2% | 3.9% | 2.9% | 5.4% | 45.2% | knn | 46.8% | 3.2% | 3.4% | 5.2% | 47.3% |
| xgbLinear | 43.6% | 4.0% | 4.4% | 5.9% | 45.2% | ranger | 44.5% | 3.4% | 0.1% | 4.7% | 46.2% |
| knn | 46.9% | 3.2% | 4.8% | 4.9% | 44.1% | rf | 44.7% | 3.7% | 2.6% | 5.5% | 44.1% |
| ranger | 43.9% | 4.0% | 4.5% | 6.0% | 44.1% | kknn | 40.9% | 4.5% | 1.7% | 6.0% | 41.9% |
| kknn | 43.0% | 4.5% | 4.8% | 5.9% | 41.9% | xgbLinear | 42.4% | 4.1% | 2.0% | 5.9% | 40.9% |

*Note: Accuracy on 10-fold cross-validation in 100 repetitions, train/test split 85:15*

The result of the second experimental run analyzing the dataset enhanced by SMOTE are shown in Table 3. The performance boost on the training set is apparent on all feature sets (for best algorithms: +34.3% likeability, +26.0% personality, +19.7% emotional valence, +24.4% basic emotions). Nevertheless, the testing set performance is drastically worse than the training set performance, revealing an extreme overfitting effect. For all feature sets, the testing set performance is approximately on the level of the unbalanced dataset equivalent – only for the emotional valence feature set, the testing set performance is distinctly lower (-7.5%) than for the unbalanced dataset. This may be explained by the low number of features which may render the estimation more sensitive to the oversampling procedure evoked by SMOTE.

## IV. DISCUSSION

The present work aimed to find the value of different feature types for predicting people's candidate choice outside an election context. The findings show that, contrary to expectations, likeability ratings do not add predictive value over the baseline. Apart from that, any psychological features – derived from personality trait ratings or from sentiment analysis – could substantially improve the prediction performance by 10-11% in holdout set accuracy. This means that emotional valence with only two features can improve prediction to a similar extent as personality with 20 features or basic emotions with 8 features. Nevertheless, the picture turned out differently when class imbalance was treated by minority oversampling (SMOTE) which distinctly changed the algorithm ranking. This result and the lack of testing set performance improvement on the oversampled dataset may, however, be specific to the dataset in the present study.

The main finding of this study is that simple likeability ratings may not explain people's choice well, but personality trait ratings or sentiments derived from people's free text answers do so to a non-negligible degree.

TABLE 3: BENCHMARKING ACCURACY RESULTS FOR ALL FEATURE SETS – SMOTE BALANCED DATASET

| psychological – likeability feature set | | | | | | psychological – personality feature set | | | | | |
|---|---|---|---|---|---|---|---|---|---|---|---|
| algorithm | Cross-valid Acc – mean | Cross-valid Acc – sd | Cross-valid Kappa – mean | Cross-valid Kappa – sd | Testingset Accuracy | algorithm | Cross-valid Acc – mean | Cross-valid Acc – sd | Cross-valid Kappa – mean | Cross-valid Kappa – sd | Testingset Accuracy |
| rf | 67.2% | 1.4% | 51.2% | 2.0% | 35.3% | gbm | 74.2% | 3.1% | 62.2% | 4.4% | 50.0% |
| ranger | 67.2% | 1.4% | 51.2% | 2.0% | 34.8% | svmRadial | 74.1% | 2.4% | 60.9% | 3.6% | 48.7% |
| xgbTree | 65.5% | 1.7% | 48.7% | 2.5% | 34.2% | rf | 74.1% | 2.9% | 62.0% | 4.2% | 44.7% |
| gbm | 64.8% | 1.7% | 47.5% | 2.6% | 33.7% | svmLinear | 72.0% | 2.9% | 59.1% | 4.1% | 43.4% |
| xgbLinear | 64.2% | 1.8% | 48.6% | 2.5% | 32.6% | kknn | 71.7% | 3.0% | 57.4% | 4.6% | 42.1% |
| nnet | 64.8% | 1.7% | 45.0% | 3.0% | 30.5% | knn | 71.4% | 3.1% | 56.4% | 4.9% | 42.1% |
| svmRadial | 65.6% | 1.3% | 45.8% | 2.2% | 29.9% | ranger | 74.2% | 2.9% | 61.7% | 4.2% | 40.8% |
| svmLinear | 63.9% | 1.4% | 42.1% | 2.5% | 25.7% | xgbLinear | 73.9% | 3.1% | 62.1% | 4.4% | 40.8% |
| knn | 59.8% | 1.7% | 33.2% | 3.2% | 18.2% | xgbTree | 74.4% | 3.1% | 62.8% | 4.5% | 40.8% |
| kknn | 60.2% | 1.7% | 37.6% | 3.0% | 17.6% | nnet | 70.5% | 2.9% | 56.3% | 4.3% | 36.8% |
| sentiments – emotional valence – feature set | | | | | | sentiments – basic emotions – feature set | | | | | |
| algorithm | Cross-valid Acc – mean | Cross-valid Acc – sd | Cross-valid Kappa – mean | Cross-valid Kappa – sd | Testingset Accuracy | algorithm | Cross-valid Acc – mean | Cross-valid Acc – sd | Cross-valid Kappa – mean | Cross-valid Kappa – sd | Testingset Accuracy |
| knn | 67.7% | 2.8% | 49.2% | 4.1% | 44.1% | nnet | 72.7% | 1.8% | 56.7% | 2.7% | 51.6% |
| kknn | 67.1% | 3.0% | 48.6% | 4.4% | 37.6% | ranger | 72.0% | 1.7% | 55.7% | 2.4% | 51.6% |
| ranger | 68.3% | 2.6% | 50.8% | 3.7% | 36.6% | rf | 71.6% | 1.8% | 55.2% | 2.6% | 51.6% |
| svmRadial | 65.9% | 2.4% | 43.7% | 4.2% | 36.6% | svmRadial | 71.3% | 1.8% | 53.6% | 2.9% | 48.4% |
| xgbLinear | 69.2% | 2.6% | 52.1% | 3.7% | 36.6% | xgbTree | 70.7% | 2.1% | 54.1% | 3.1% | 48.4% |
| rf | 68.0% | 2.7% | 50.7% | 3.8% | 35.5% | knn | 71.2% | 2.1% | 54.4% | 3.3% | 47.3% |
| gbm | 68.7% | 2.5% | 50.9% | 3.8% | 34.4% | gbm | 70.5% | 2.2% | 53.4% | 3.3% | 46.2% |
| nnet | 65.7% | 2.5% | 42.7% | 4.4% | 34.4% | xgbLinear | 69.6% | 2.3% | 53.0% | 3.3% | 46.2% |
| xgbTree | 69.2% | 2.5% | 51.8% | 3.7% | 34.4% | svmLinear | 70.1% | 1.9% | 51.2% | 3.2% | 45.2% |
| svmLinear | 62.2% | 2.4% | 32.9% | 4.9% | 26.9% | kknn | 69.3% | 2.6% | 52.1% | 4.0% | 41.9% |

Note: Accuracy on 10-fold cross-validation in 100 repetitions, train/test split 85:15

ACKNOWLEDGMENT

The author would like to thank Takeshi Teshima, Tokyo University, for the interesting discussion on an earlier draft of this paper. This research was supported by the Yonsei University Faculty Research Fund of 2019-22-0199.